\documentclass{article}
\usepackage{graphicx} 
\usepackage{cite}
\usepackage{amsmath,amssymb,amsfonts}
\usepackage{algorithmic}
\usepackage{graphicx}
\usepackage{textcomp}
\usepackage{lineno,hyperref}
\modulolinenumbers[5]
\usepackage{multirow}
\usepackage{epstopdf}
\usepackage{tabularx}
\usepackage{xcolor}
\usepackage{placeins}
\usepackage{setspace}
\usepackage{booktabs}
\usepackage{graphicx}
\usepackage{longtable}
\usepackage{amsmath}
\usepackage{amstext}
\title{Investigating Metal Dopants for Lowering the Contact Resistance of Top Gold Contacted Monolayer MoS$_2$}
\author{Saurabh Kharwar \thanks{MDCL lab, Department of electrical engineering, Indian Institute of Technology, Gandhinagar-382355, India,  saurabh.ec17@nitp.ac.in}, Soham Sinha, and Tarun Kumar Agarwal}

\begin{document}

\maketitle
\begin{abstract}
The interface properties between gold (Au) contacts and molybdenum disulfide (MoS$_2$) are critical for optimizing the performance of semiconductor devices. This study investigates the impact of metal dopants (D) on the transport properties of MoS$_2$ devices with top Au contacts, aiming to reduce contact resistance and enhance device performance. Using density functional theory (DFT) and non-equilibrium Green’s function (NEGF)-based first-principles calculations, we examine the structural, electronic, and quantum transport properties of Au-contacted, metal-doped MoS$_2$. Our results indicate that Cd, Re, and Ru dopants significantly improve the structural stability and electronic properties of MoS$_2$. Specifically, formation energy calculations show that Cd and Re are stable at hollow sites, while Ru prefers bond sites. Remarkably, Au-Ru-MoS$_2$-based device exhibits tunnel resistance (R$_T$) up to 4.82 $\Omega$.$\mu$m. Furthermore, a dual-gated Au-Ru-MoS$_2$ field effect transistor (FET) demonstrates an impressive I$_{on}$/I$_{off}$ ratio of 10$^{8}$ at V$_{gs}$ of 2 V, highlighting its potential for nano-switching applications.
\end{abstract}
\textbf{Keywords-}Metal-Semiconductor interface, Electronic structure, Transport properties, Device resistance.
\section{Introduction}
\label{sec:introduction}
In the rapidly evolving field of semiconductor device research, a primary goal is to develop efficient and transparent ohmic contacts between semiconductors and metal electrodes. These contacts are crucial for effectively injecting charge carriers into the conduction channel, a necessity for high-performance electronic devices \cite{zheng2021ohmic}. As electronic technologies progress beyond silicon, there is a growing need for new channel materials and ultra-low-resistance contacts \cite{chhowalla2016two,allain2015electrical}. Atomically thin two-dimensional (2D) semiconductors hold great promise for enhancing high-performance electronics \cite{chhowalla2016two,akinwande2019graphene}. The introduction of 2D materials, characterized by their atomically flat surfaces free of dangling bonds, presents an exciting opportunity for creating these essential contacts.\\
\indent Despite its potential, 2D devices often have high Schottky barriers and frequently do not follow the Schottky-Mott rule due to interfacial phenomena such Fermi-level pinning \cite{zheng2021ohmic}. Nonetheless, these materials address significant problems at the metal-semiconductor (M-S) interface, principally due to metal-induced gap states (MIGS), which cause high contact resistance and limit current delivery capabilities \cite{allain2015electrical,louie1976electronic,sotthewes2019universal,tung2014physics,razavieh2019challenges}. M-S contacts are critical in modern electronics, particularly as device sizes continue to reduce \cite{li2023approaching}. Ideally, an M-S junction should achieve a fundamental quantum limit in junction contact resistance, assuming ballistic transport of the conducting mode.
The situation is more complex for van der Waals (vdW) materials, such as transition metal dichalcogenides (TMDs), due to their dangling-bond-free surfaces. The electrical contact resistance at this interface continues to be a key issue, limiting semiconductor device growth and performance \cite{razavieh2019challenges}. This resistance is primarily generated by the Schottky barrier \cite{tung2014physics}, which derives from the energy difference between the metal work function and the semiconductor electron affinity, as well as MIGS, which results in Fermi-level pinning \cite{louie1976electronic,sotthewes2019universal}. When a semiconductor comes into close contact with a metal surface, the metal's extended wavefunction perturbs the semiconductor environment, generating rehybridizations of the semiconductor's wavefunctions.
Several approaches have been developed to improve contact resistance. One method is to reduce the Schottky barrier width by severely doping the semiconductor. This process, known as doping, inserts impurities into the semiconductor, altering its characteristics and increasing the tunneling current's strength over the thermionic emission current at the Schottky barrier. However, this approach presents technological obstacles, particularly for two-dimensional materials \cite{liu2016van}. Another approach is to create a gap between the metal and the semiconductor to dissociate the interaction. This can be accomplished by introducing a thin dielectric layer, a molecular layer, or a van der Waals gap to the interface, lowering resistance \cite{vilan2000molecular,cui2017low}.  
Despite showing promise, the increased metal-semiconductor distance in this method frequently leads to a non-negligible tunneling barrier \cite{liu2016van,liu2018approaching}. Thus, in monolayer TMD transistors, these approaches generally result in either large Schottky barriers (between 100 and 400 meV) or interface tunneling barriers thicker than 1 nm \cite{kobayashi2009fermi,wang2019van,cui2017low,kim2017fermi}. Contact resistance values are many orders of magnitude greater than the quantum limit due to the extra tunnel barrier introduced by the vdW gap, which further reduces charge injection \cite{landauer1957spatial}.\\
\indent There are several innovative techniques being used to attempt to bridge this ``contact gap". These comprise low-work-function metals \cite{das2013high}, ultra-high vacuum evaporation \cite{english2016improved}, doping \cite{fang2013degenerate,mcclellan2021high}, tunneling contacts \cite{cui2017low,wang2016high}, edge contacts \cite{jain2019one,cheng2019immunity}, and, more recently, semi-metal contacts \cite{shen2021ultralow,chou2021antimony,o2021advancing,chou2020high}. By improving charge injection efficiency and lowering contact resistance, these techniques seek to further push the limits of 2D semiconductor device performance. As of right now, the state-of-the-art contact resistance is still about 1 K$\Omega$.$\mu$m, which is significantly greater than metal-Si contacts \cite{chhowalla2016two}.
In order to achieve next-generation electronics, channel material thickness must be scaled down to the two-dimensional limit while keeping ultra-low contact resistance \cite{akinwande2019graphene,ahmed2020introducing}. Transistor scaling is supported by TMDs all the way to the conclusion of the technology roadmap. Device performance remains limited by contact restrictions despite major efforts \cite{wang2022making,das2021transistors}. Due to the intrinsic vdWs gap, the contact resistance in TMD devices has not exceeded that of covalently bonded M-S junctions, and the most sophisticated contact technologies are experiencing stability problems \cite{wang2022making,shen2021ultralow}. 
\\
\indent Motivated by the need to address the challenges posed by contact resistance in M-S interface, this study focuses on investigating the electronic and transport properties of doped MoS$_2$ with Au contact. MoS$_2$, a two-dimensional semiconductor, has garnered considerable attention due to its unique electronic properties and potential applications in nanoelectronic devices. Doping MoS$_2$ with foreign atoms such as Cd, Re, and Ru offers a versatile platform for tailoring its electronic structure and enhancing device performance. By studying the effect of Au contact on doped MoS$_2$, we aim to gain insights into the role of metal-semiconductor interfaces in modulating resistance and elucidate the underlying mechanisms governing charge transport across the vdW gap. This research holds promise for advancing our understanding of metal-semiconductor heterostructures and facilitating the development of high-performance electronic devices for diverse applications.\\

\section{Computation Method}
\begin{figure}[htbp]
	\centering
	\includegraphics[height=10cm, width=9cm]{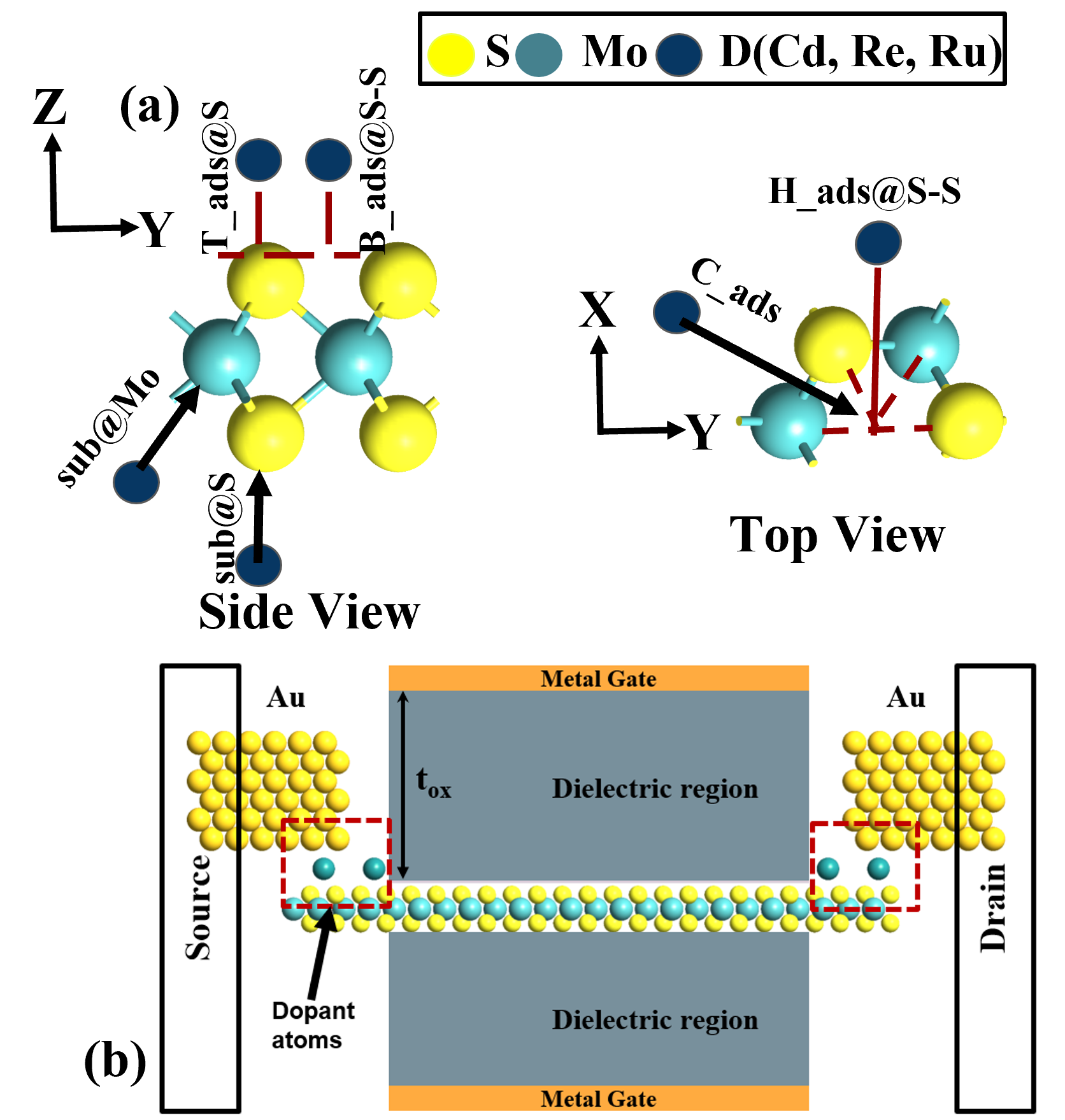}
	\caption{(a) Side view and top view of the atomic sites in a monolayer MoS$_2$, illustrating both substitutional and adsorbed foreign D atoms within a 1x1 cell. Subtitutionally doped at Mo and S positions are referred as sub@Mo and sub@S respectively. Adsorbed site at top, bridge, hollow, and center sites are referred as ads@T, ads@B, ads@H, and ads@C respectively. (b) MoS$_2$-based FET showing integration with Au contacts.}
	\label{fig1}
\end{figure}

In this work, we have studied the effect of Cd, Re, and Ru dopant metal (D) atoms in MoS$_2$ (D-MoS$_2$) on its electronic transport properties contacted with Au using a multi-scale framework. The multi-scale framework combines the Density functional theory (DFT) calculations for the D-MoS$_2$ system, Au-D-MoS$_2$ system and electron transport calculations of the top Au-D-MoS$_2$ system with open boundary conditions.\\ 
\indent The atomic configurations and field effect transistor (FET) device of D-MoS$_2$ are shown in Figure \ref{fig1}. Figure \ref{fig1}a shows 1x1 cell with 2 Mo and 4 S atoms which is used to study the most stable position of the metal atoms. To examine the most stable site of metal atom, substitution doped and suface adsorbed sites are choosed. Subtitutionally doped at Mo and S postions are referred as sub@Mo and sub@S respectively. Further, adsorbed site at top, bridge, hollow, and center sites are referred as ads@T, ads@B, ads@H, and ads@C respectively  \\
\indent Density functional theory (DFT) combined with non-equilibrium Green's function (NEGF) method are deployed to carry out the first-principles computations implemented in the QuantumATK tool \cite{brandbyge2002density}. The energy cutoff for the plane waves is 150 Ry and the Brillouin zone was sampled using a $5\times5\times1$ Monkhorst-pack grid. The Generalized Gradient Approximation (GGA) approach is used for the exchange correlation. A double-zeta-polarized (DZP) basis set is used. 
A vacuum padding of 40 {\AA} is used in non-periodic direction to avoid the periodic image replicas. The force tolerance of 0.01 eV/{\AA} is used as the convergence criterion for relaxation structures. The stability of the considered structures are calculated via formation energy (E$ _{form} $) \cite{kharwar2022dft},
\begin{equation}  
	E_{form}= E_{D-MoS_2} - E_{MoS_2} - \mu_M + \mu_Y
\end{equation} 	

Here, $E_{D-MoS_2}$, $E_{MoS_2}$ and $\mu_{M/Y}$ represents the total energy of the D-MoS$_2$, MoS$_2$ structure, and chemical potential of M-metal/removed-atoms respectively. The Landauer-B$\ddot{u}$ttiker formalism is used to compute the current-voltage (I-V) characteristics of the studied devices \cite{brandbyge2002density,kharwar2021first,kharwar2022dft}, 
\begin{equation}  
	G_C(E)= [{EI-H-\Sigma_{L}^{r}(E)-\Sigma_{R}^{r}(E)}]^{-1}
	\label{eq2}
\end{equation}
\begin{equation}  
	T(E)=T_r[\tau_R(E)G_C(E)\tau_L(E)G_C^+(E)]
	\label{eq3}
\end{equation}
\begin{equation}  
	I(V_b)=\frac{2e^2}{h}\int_{\varepsilon_L}^{\varepsilon_R}T(E,V_b)[F(E-\varepsilon_L)-F(E-\varepsilon_R)]dE
\end{equation} 
Where the variables G$_C$(E) and G$_C^+$(E) represent the retarded Green's function, advanced Green's function of channel region. 
T(E) and I(V$_b$) represent the transmission coefficient and current at the applied bias voltage (V$_b$). The Hamiltonian and identity matrices for the retarded Green's function are denoted by $H$ and $I$, respectively. 
The self-energies, coupling coefficients, and electrochemical potential of the left/right electrodes are denoted by $\Sigma_{L/R}^{r}$, $\tau_{(L/R)}$, and $\varepsilon_{L/R}$, respectively. \\
\section{Results and Discussion}
\subsection{Cd/Re/Ru doped \texorpdfstring{MoS$_2$}- super cell}
\begin{figure}[htbp]
	\centering
	\includegraphics[height=6cm, width=8cm]{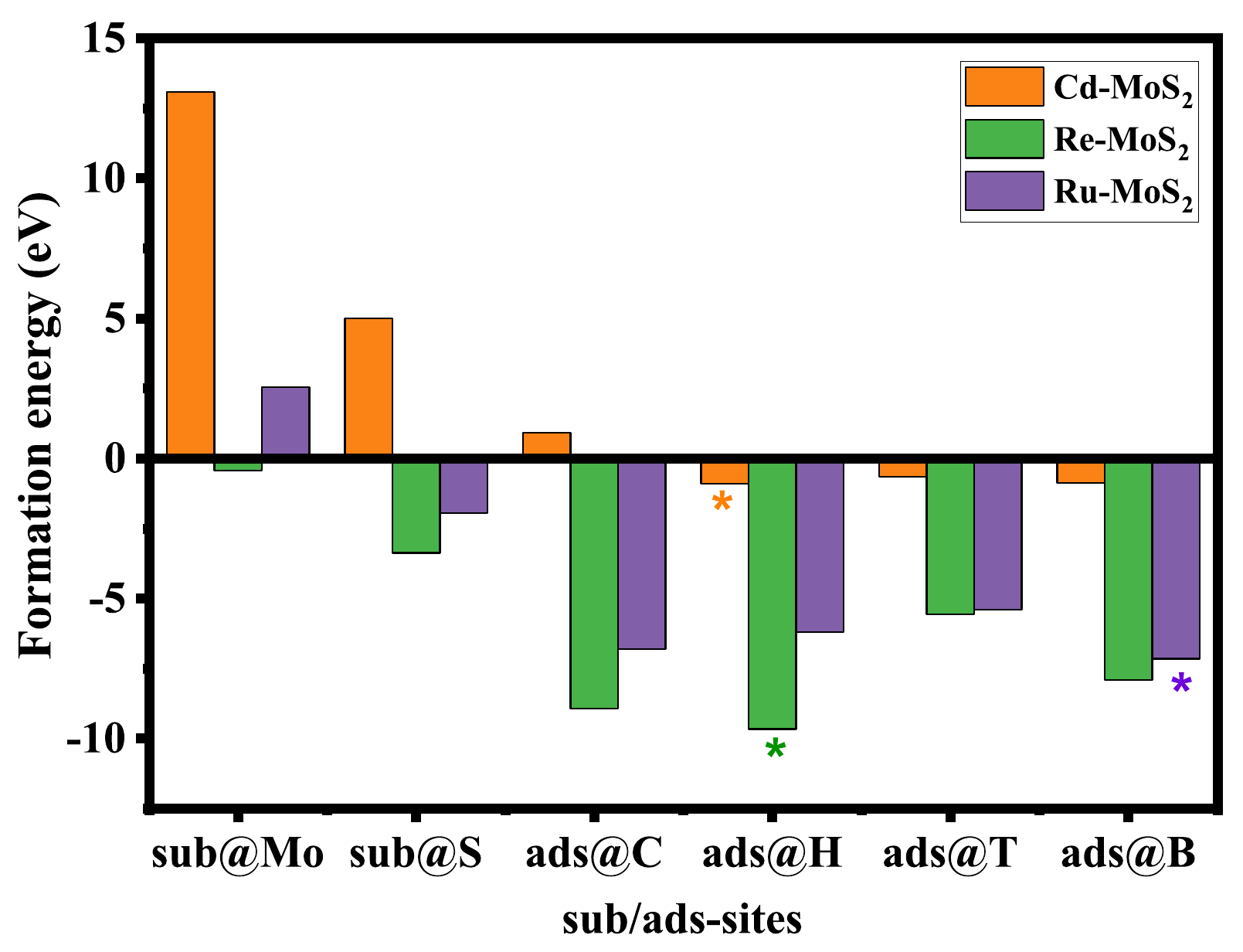}
	\caption{Formation energy for all three D atoms at different sites for 1x1 cell. Star symbols are used to indicate the most stable sites of dopant atoms. }
	\label{fig2}
\end{figure}

\begin{figure}[htbp]
	\centering
	\includegraphics[height=9cm, width=11cm]{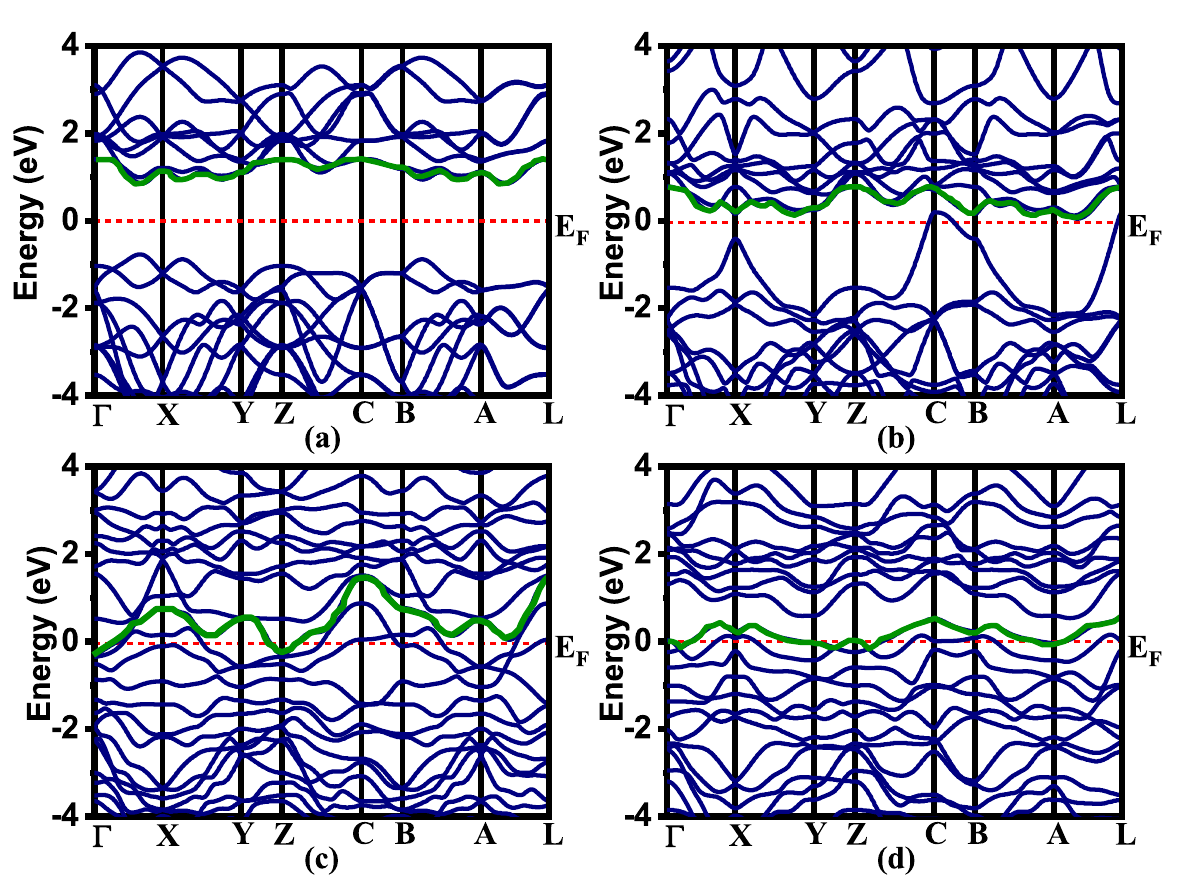}
	\caption{Bandstructure of (a) MoS$_2$, (b) Cd-MoS$_2$, (c) Re-MoS$_2$, and (d) Ru-MoS$_2$ respectively. Fermi level is represented with red dotted line at 0 eV. The green color bands are used to calculate effective masses. }
	\label{fig4}
\end{figure}

\begin{figure}[htbp]
	\centering
	\includegraphics[height=12cm, width=8cm]{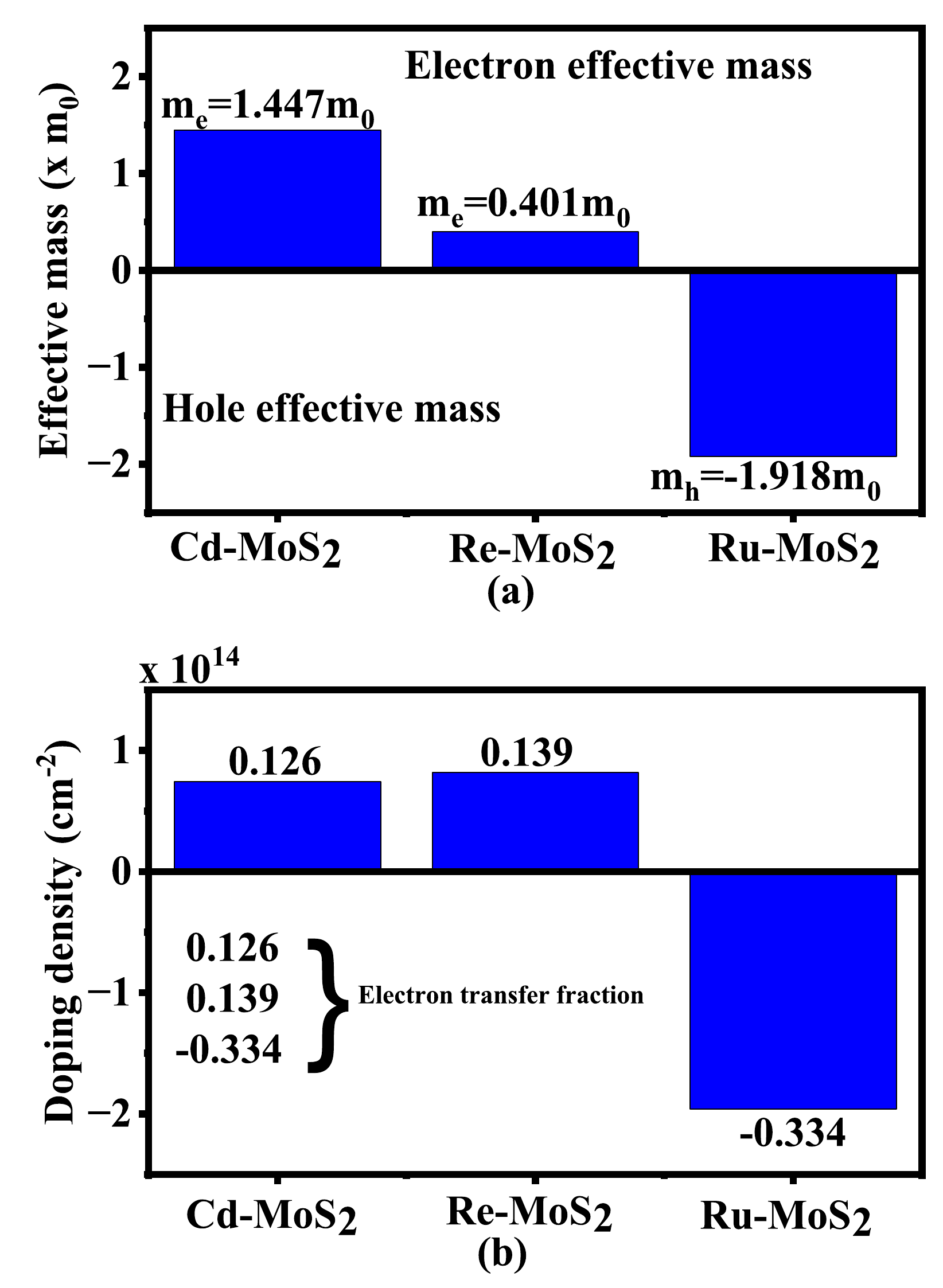}
	\caption{(a) The effective mass and (b) doping density of Cd-MoS$_2$, Re-MoS$_2$, and Ru-MoS$_2$ respectively.  }
	\label{fig3}
\end{figure}

\begin{figure}[htbp]
	\centering
	\includegraphics[height=15cm, width=12cm]{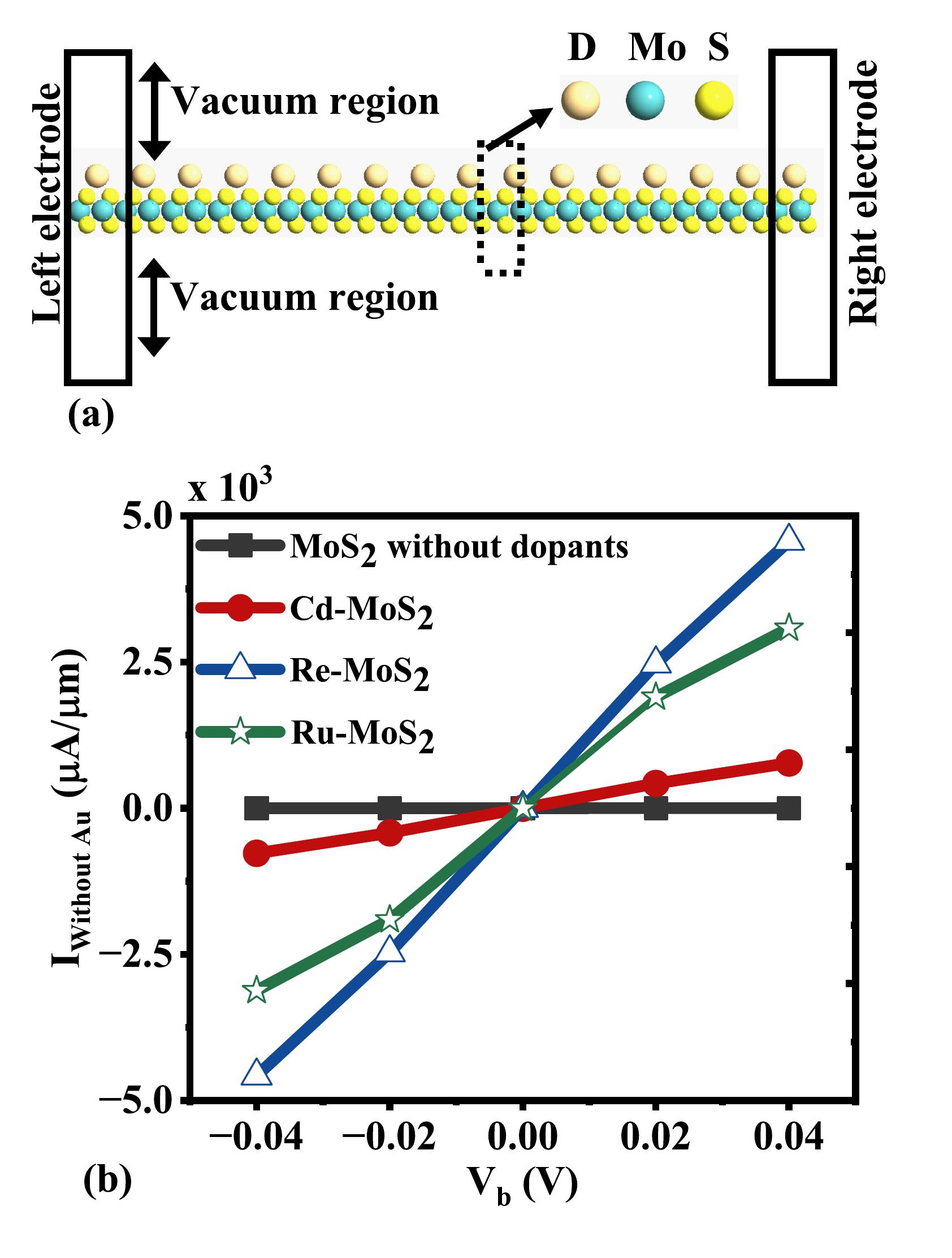}
	\caption{(a) Two-terminal device configuration without Au contact and (b) I-V characteristics of D-MoS$_2$ devices respectively.  }
	\label{fig7}
\end{figure}

To investigate the most stable position of the dopant atom in the pristine MoS$_2$ cell, we find the $E_{form}$ of the dopant atoms at multiple sites as shown in the representative Figure \ref{fig1}a. As shown in Figure \ref{fig1}a, the D atoms have been investigated for various substitutional and adsorbed sites. Figure \ref{fig2} exhibits ads@H (adsorption of the D atom at a hollow site) as the most stable site due to its lowest formation energy for three different D atoms.\\
\indent The calculation of bandstructures for MoS$_2$ and MoS$_2$ doped with Cd, Re, and Ru reveals intriguing insights into the electronic properties of these materials as shown in Figure \ref{fig4}. MoS$_2$ exhibits a typical semiconductor band structure characterized by a bandgap between the valence and conduction bands. However, upon doping with Cd, Re, or Ru atoms, additional electronic states emerge near the Fermi level (E$_F$), leading to a metallic nature. This phenomenon can be attributed to the introduction of dopant-induced electronic states within the bandgap, which effectively alter the electronic structure of the material. The presence of these additional bands at the E$_F$ facilitates the delocalization of charge carriers, resulting in enhanced electrical conductivity and the manifestation of metallic behavior. This observation underscores the significant impact of dopants on the electronic properties of MoS$_2$ and highlights the potential for tailoring its conductivity characteristics for various electronic applications.\\
\indent To understand the interaction and type of doping by the D-atoms for adsorption site, we calculate the effective mass and doping density by electron transfer to the MoS$_2$ cell using Mulliken Population Analysis, shown in Figure \ref{fig3}. The effective mass of Cd-MoS$_2$ and Re-MoS$_2$ are calculated in transport direction i.e. y-direction from the conduction band minima (CBM) while valence band maxima (VBM) is used for Ru-MoS$_2$ as shown by green color band in Figure \ref{fig4}. Moreover, the doping densities along with charge transfer values at the top of bar are shown in Figure \ref{fig3} (b). Adsorption of Ru results in lesser electron charge (i.e. p-doping) while adsorption of Cd, and Re result in increased electron charge (i.e. n-doping). Subsequently, the doping densities of each structure can be calculated by dividing electron transfer to MoS$_2$ and cell area. Figure \ref{fig3} (b) reports 7.41E+13, 8.18E+13, and -1.96E+14  per cm$^2$ doping densities for Cd, Re, and Ru-MoS$_2$, respectively. It is to be noted that the calculated doping densities are higher than typical experimentally reported doping values to MoS$_2$ \cite{english2016improved} due to the smaller MoS$_2$ cell size. This cell area is chosen to investigate the lowest limit of contact resistance, as it is understood that contact resistance increases with reduced doping concentration.\\
\indent Moreover, We have also studied the effect of D-atom adsorption on the MoS$_2$-based two-terminal devices as shown in Figure \ref{fig7}. The length of the channel region is about 77 {\AA}. The current-voltage characteristics of Cd-MoS$_2$, Re-MoS$_2$, and Ru-MoS$_2$ devices demonstrate the direct impact of effective mass on the electrical performance of semiconductor devices. Re-MoS$_2$, with the lowest effective mass, exhibits the highest current due to its superior carrier velocity. Ru-MoS$_2$, despite having a negative effective mass, still shows significant current, likely due to complex band structure interactions. Conversely, Cd-MoS$_2$, with the highest effective mass, has the lowest current, confirming the theoretical relationship between effective mass, carrier velocity, and current.\\

\subsection{Au contacted Cd/Re/Ru doped \texorpdfstring{MoS$_2$}- 
 interface }
\begin{figure}[htbp]
	\centering
	\includegraphics[height=4.5cm, width=9cm]{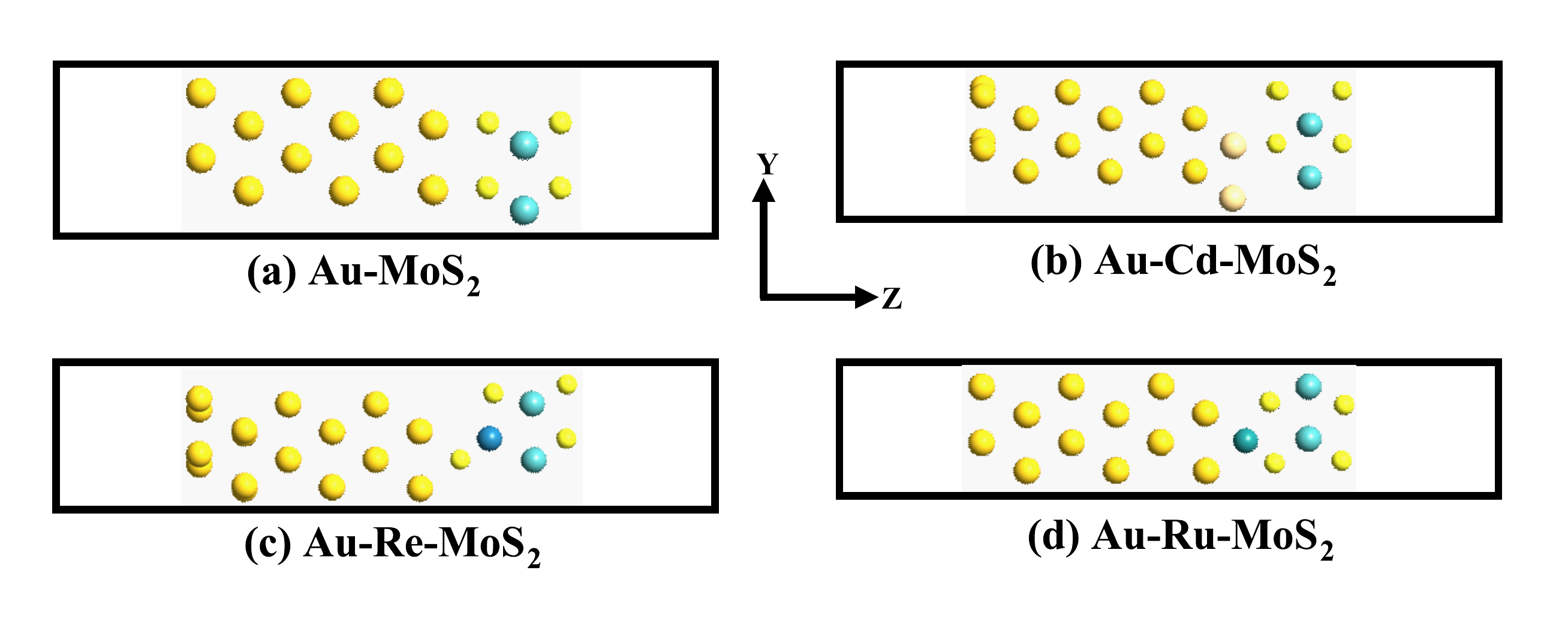}\\
	\includegraphics[height=12cm, width=8cm]{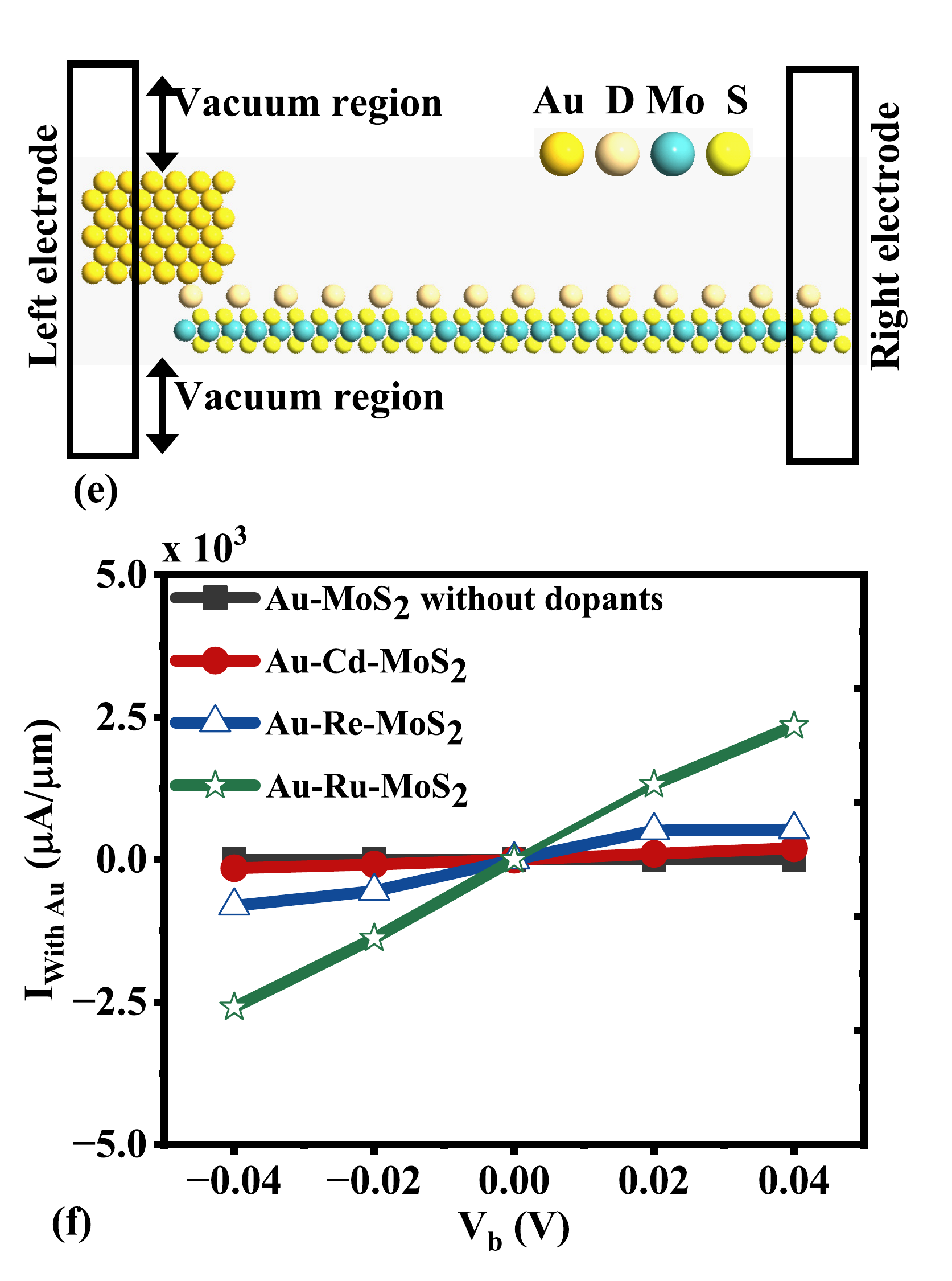}
	\caption{(a-d) Au-D-MoS$_2$ interfaces, (e) Au-D-MoS$_2$ device configuration, and (f) I-V characteristics of Au-D-MoS$_2$ devices respectively.}
	\label{interface}
\end{figure}

\begin{figure*}[htbp]
	\centering
 \includegraphics[height=5cm, width=6cm]{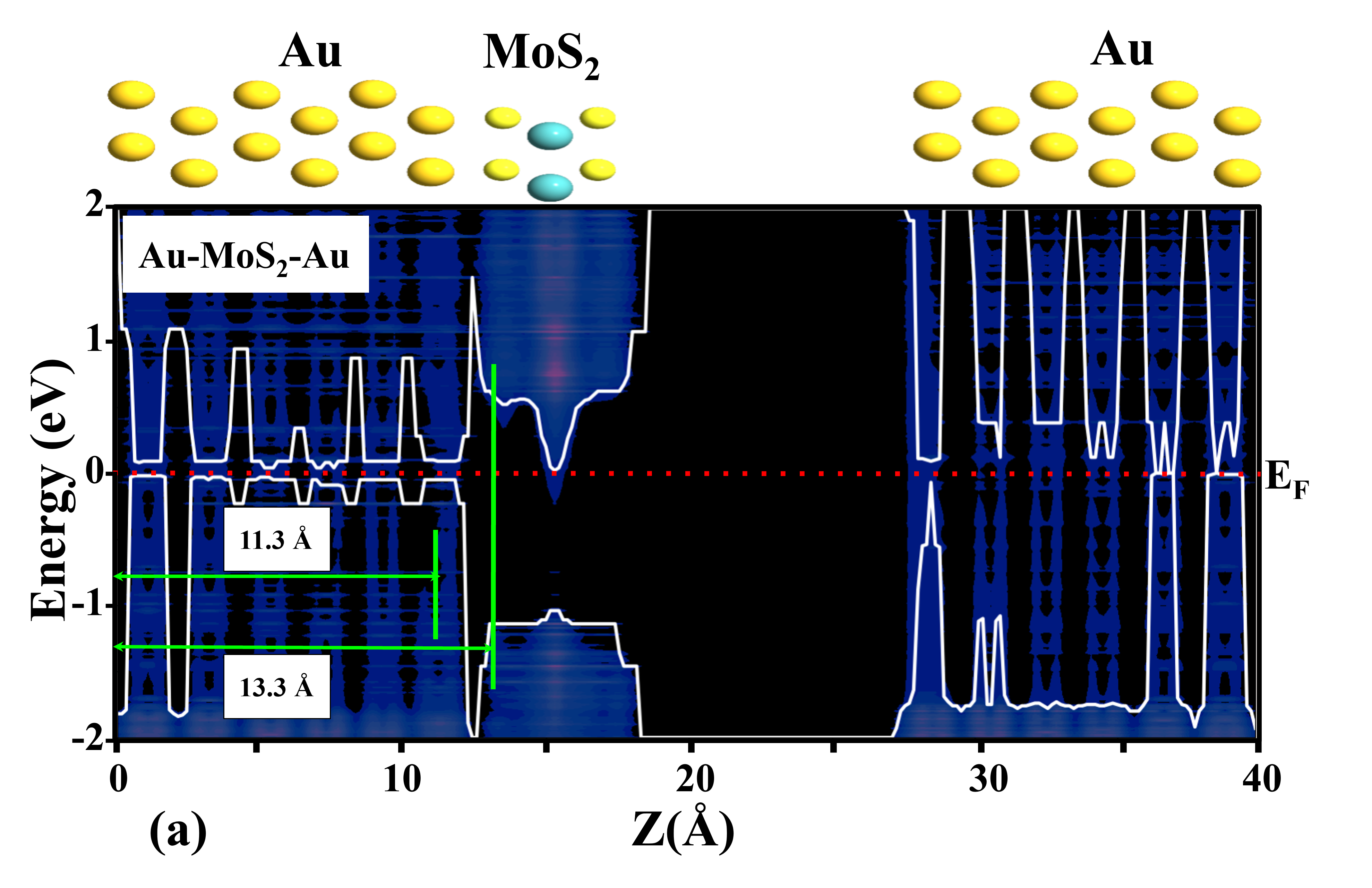}
 \includegraphics[height=5cm, width=6cm]{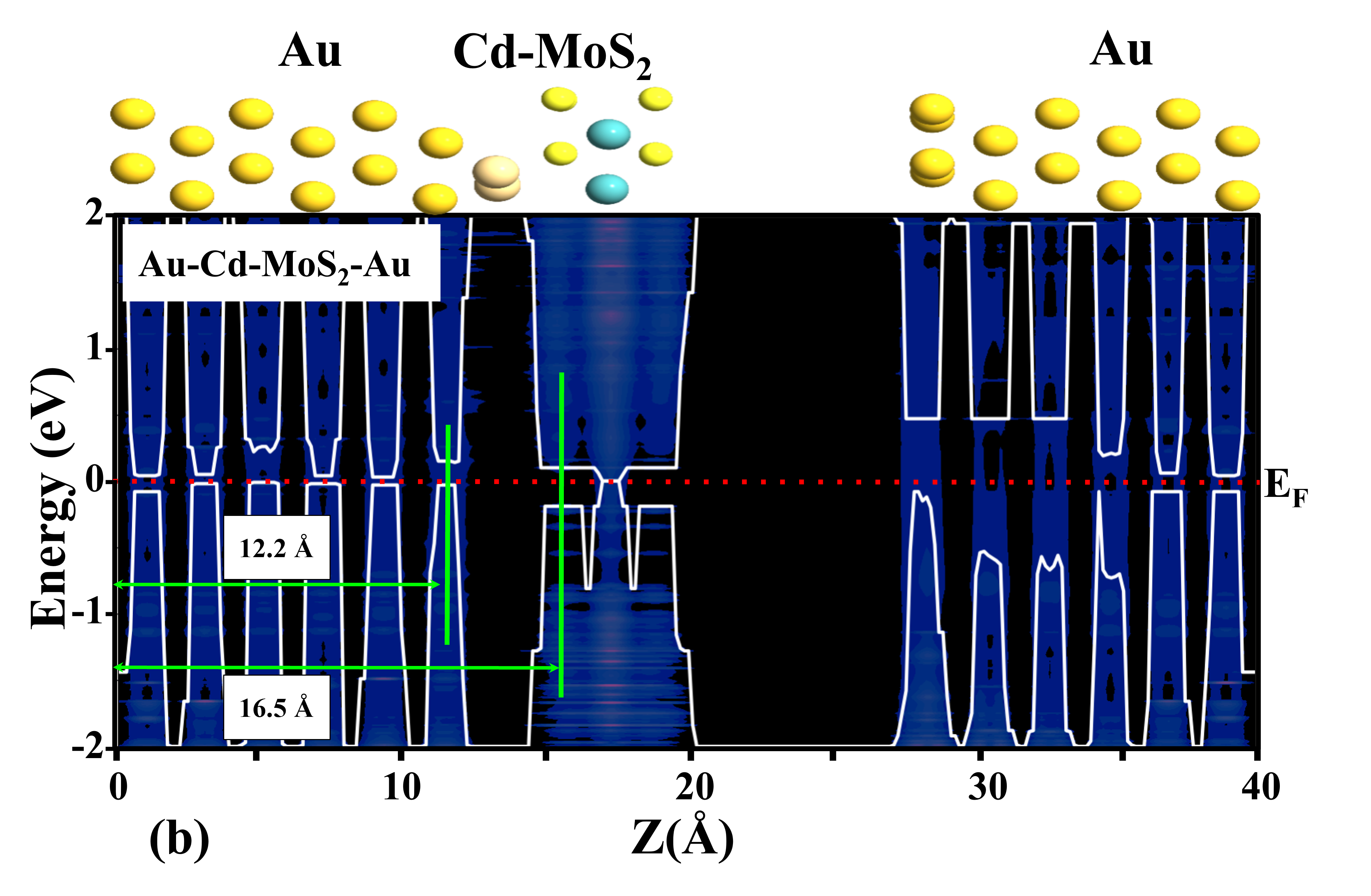}
 \includegraphics[height=5cm, width=6cm]{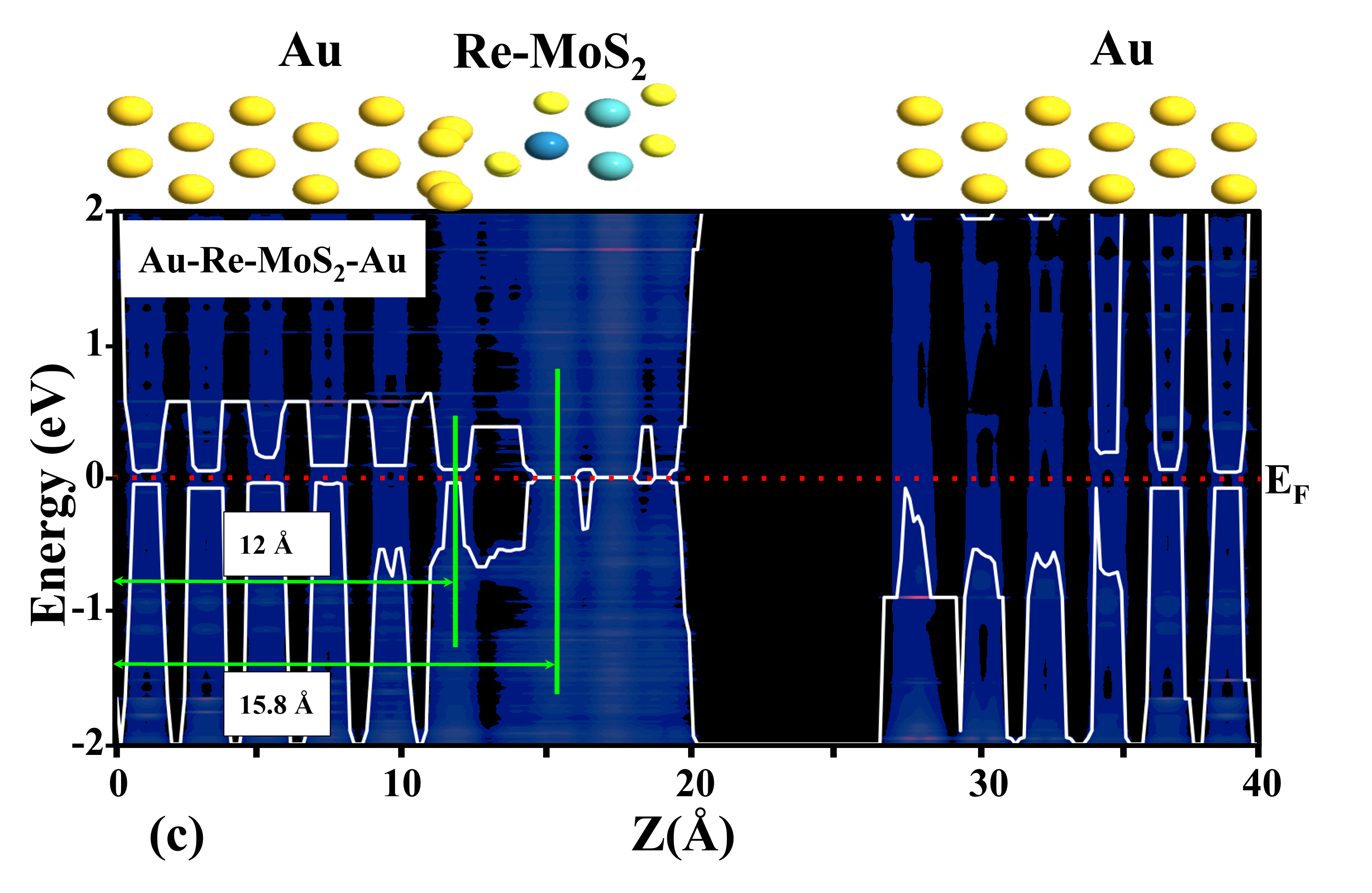}
 \includegraphics[height=5cm, width=6cm]{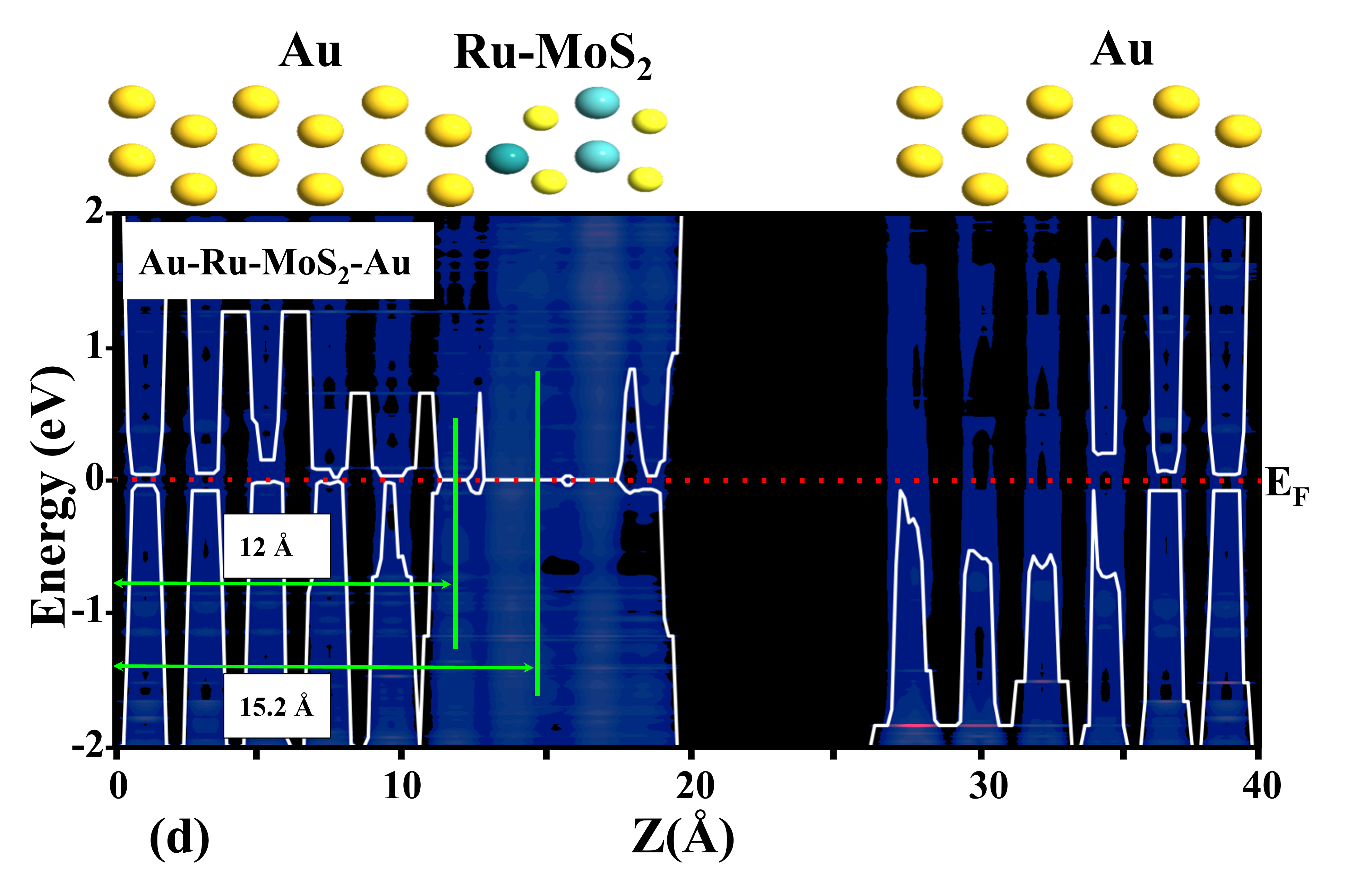}
	\caption{PLDOS of verticle Au-MoS$_2$-Au devices with (a) Au-MoS$_2$-Au, (b) Au-Cd-MoS$_2$-Au, (c) Au-Re-MoS$_2$-Au, and (d) Au-Ru-MoS$_2$-Au respectively. }
	\label{PLDOS of verticle device}
\end{figure*}

Next, we attach a Au contact to the Cd/Re/Ru doped MoS$_2$ as shown in Figure~\ref{interface}. Mulliken population Analysis is used to calculated the charge transfer per atom on MoS$_2$. The obtained charge transfer on MoS$_2$ for Au-Cd-MoS$_2$, Au-Re-MoS$_2$, and Au-Ru-MoS$_2$ are 0.022e, 0.008e, and -0.051e respectively. It shows that the magnitude of electron fraction to MoS$_2$ has increased significantly for Cd-MoS$_2$ system, while electron fraction on MoS$_2$ has decreased in Re-MoS$_2$ and Ru-MoS$_2$ systems after Au contact.\\
\indent Moreover, we investigated the behavior of two-terminal devices with one side contacted by Au electrodes while the other remained without Au as shown in Figure \ref{interface}(e). The length of the channel region is about 80 {\AA}. 
The I-V characteristics of four devices (Au-MoS$_2$, Au-Cd-MoS$_2$, Au-Re-MoS$_2$, and Au-Ru-MoS$_2$) are analyzed, revealing distinct trends as shown in Figure \ref{interface}(f). Interestingly, initially, the Re-doped MoS$_2$ device without Au contact displayed ohmic behavior compared to Cd-MoS$_2$ and Ru-MoS$_2$ devices. Upon introducing Au contact, the doped-MoS$_2$ device exhibited ohmic behavior for Au-Ru-MoS$_2$.\\
\indent Furthermore, we have studied the Au-MoS$_2$-Au-based vertical two-terminal device and calculated the Projected Local Density of States (PLDOS) (see Figure \ref{PLDOS of verticle device}) to unravel the underlying physical mechanisms within the van der Waals (vdW) gap. In these devices, the Au contact are attached at both side of D-MoS$_2$. The right side Au contact is kept to be at a distance of 10 {\AA} because interest is in only study the effect of left Au on MoS$_2$. Notably, in pristine Au-MoS$_2$-Au configurations, we observed a finite gap within the vdW region. However, when dopants such as Cd, Re, and Ru were introduced, distinct electronic behaviors emerged. In the Au-Cd-MoS$_2$-Au device, while a gap persisted within the vdW gap, energy states associated with Cd atoms were detected near the Au layer. Intriguingly, in Au-Re-MoS$_2$-Au and Au-Ru-MoS$_2$-Au configurations, no discernible gap was observed within the vdW region. 
Furthermore, we quantified the changes in I-V characteristics before (see Figure \ref{fig7}b) and after Au contact (see Figure \ref{interface}(f)), revealing minimal alterations for Cd and Re-doped MoS$_2$ devices, whereas the Ru-MoS$_2$ device showed significant changes. These findings suggest that the presence of Au contact influences the electrical behavior of doped MoS$_2$ devices, with Ru-doped MoS$_2$ exhibiting notable sensitivity to Au contact.\\
Furthermore, the average tunnel resistance (R$_T$) are calculated by analyzing the difference in average resistance between a D-MoS$_2$ device without Au contact (R$_{D-MoS_2}$) and an Au-D-MoS$_2$ device (R$_{Au-D-MoS_2}$) from -0.04 V  to 0.04 V. 
The calculated (R$_T$) of the considered devices are shown in Table \ref{tab:1}. The R$_T$ of Au-Ru-MoS$_2$ device is found to be very low 4.82 $\Omega$.$\mu$m. Furthermore, the ballistic limit of contact resistance (R$_C$) are calculated using \cite{jena2014intimate},
\begin{equation}  
R_C = \frac{0.026\times1000}{\sqrt {n_{2D}}} \Omega.{\mu}m, \hspace{10pt} where \hspace{10pt} {n_{2D} = \frac{N_{2D}}{10^{13}cm^{-2}}}
	\label{eq5}
\end{equation}
where, $N_{2D}$ is ${E_{F}}/{Area}$ and $E_{F}$ is sum of electron fraction transfer to MoS$_2$ in D-MoS$_2$ system and Au-D-MoS$_2$ system respectively.\\
The obtained values of R$_C$ are 11.88 $\Omega$.$\mu$m, 11.91 $\Omega$.$\mu$m, and 7.37 $\Omega$.$\mu$m for Au-Cd-MoS$_2$, Au-Re-MoS$_2$, and Au-Ru-MoS$_2$ device respectively. The investigation into the I-V characteristics of D-MoS$_2$ devices without Au contacts revealed that Re-MoS$_2$ exhibited the highest current due to its lower effective mass, which enhances carrier velocity and mobility. In a subsequent analysis of Au-D-MoS$_2$ devices, it was found that Au-Ru-MoS$_2$ showed the highest current values and the lowest R$_T$, indicating an optimized tunneling interface. Calculations of R$_C$ for Au-D-MoS$_2$ devices demonstrated that Au-Ru-MoS$_2$ approached the theoretical R$_C$, signifying minimal resistive losses. Qualitatively, the absence of states in the PLDOS within the van der Waals (vdW) gap (see Figure \ref{PLDOS of verticle device}) suggests complex interactions beyond mere metal doping, impacting potential barriers and device performance. Quantitatively, while Ru-doped devices exhibited higher ballistic currents and a closer approach to ballistic limits, Cd and Re-doped devices did not, despite higher doping levels. This discrepancy underscores the need for sophisticated models that consider the vdW gap's intricate physics, as N$_{2D}$ alone is insufficient for predicting ballistic performance. 

\subsection{Au contacted \texorpdfstring{Ru-MoS$_2$-based}- FET}
\begin{figure*}[htbp]
	\centering
	\includegraphics[height=10cm, width=16cm]{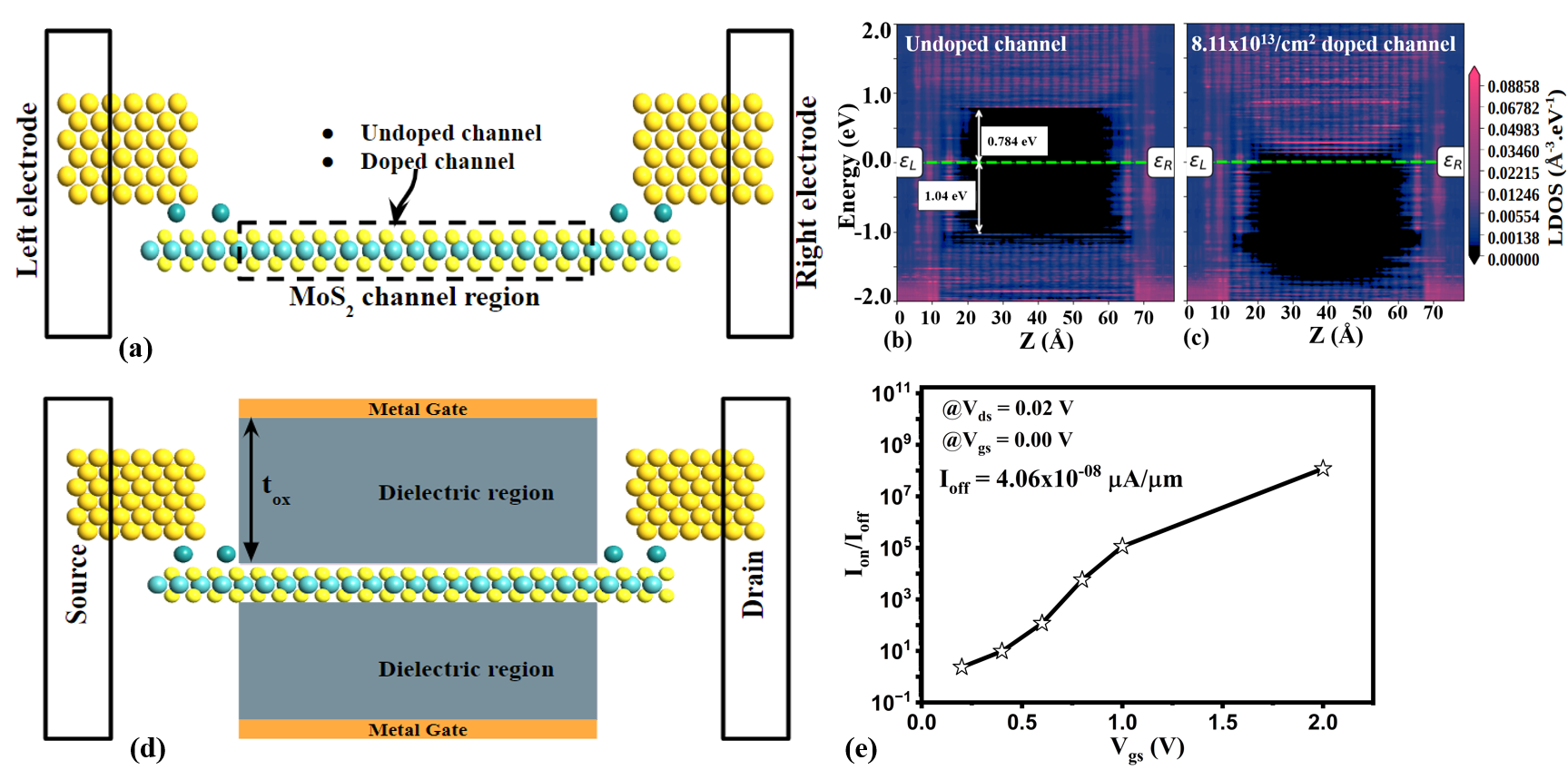}
	\caption{(a) Au-Ru-MoS2 device, (b-c) PLDOS, (d)  device configuration of double gated FET and (e) I$_{on}$/I$_{off}$ with V$_{gs}$ variations respectively. }
	\label{FET}
\end{figure*}

\indent Here, we analyzed the I-V characteristics of Au-Ru-MoS$_2$-based field-effect transitor (FET) because of low R$_T$ and identified elevated current levels attributed to the metallic nature of the channel material operating in depletion mode. To address this issue, we shifted our focus to a two-terminal device featuring an undoped MoS$_2$ channel region and Au-Ru-MoS$_2$ contacts at both terminals (as depicted in Figure \ref{FET}(a)).  The length of the channel region is about 80 {\AA}. \\
\indent We evaluated the I-V characteristics of both undoped and electrostatically doped channels. Initially, the recorded current values were notably low. To enhance conductivity, we introduced charge doping at a concentration of $8.11\times10^{13}$/cm$^2$ within the channel region surrounding MoS$_2$ atoms. This doping process effectively increased the device's conductivity. The computed PLDOS for both undoped and doped devices revealed a reduction in vertical barrier height, indicating a transition to the ON state (Figure \ref{FET}(b-c)).\\	
\indent Additionally, we examined the I-V characteristics of dual-gate transistors to assess the effect of gate voltages (V$_{gs}$). The dual-gate device configuration of Au-Ru-MoS$_2$ FET with an oxide thickness (t$_{ox}$) of 20 {\AA} and dielectric constant of 19 $\epsilon$$_0$ is illustrated in Figure \ref{FET} (d). The I$_{off}$ of the FET is observed to be $4.06 \times 10^{-08} {\mu}A/{\mu}m$. Remarkably, the FET exhibited ON state behavior at V$_{gs}$ from 2 V, with high  I$_{on}$/I$_{off}$ observed up to $10^8$ at V$_{gs}$ of 2 V (Figure \ref{FET} (e)). These findings suggest that the Ru ads@H site of MoS$_2$ holds promise as a strategy for improving device performance, indicated by the achieved low R$_T$ and high  I$_{on}$/I$_{off}$ ratios.
\section{Conclusion}
In conclusion, this study demonstrates that doping MoS$_2$ with Cd, Re, and Ru significantly enhances its electronic transport properties when interfaced with Au contacts. Using a multi-scale framework combining DFT and NEGF calculations, we identified that Cd and Re are stable at hollow sites, while Ru prefers bond sites. These dopants introduce additional electronic states near the Fermi level, leading to a dramatic reduction in device tunnel resistance by up to 4.82 $\Omega$.$\mu$m. Notably, the Au-Ru-MoS$_2$ FET exhibited an exceptional  I$_{on}$/I$_{off}$ ratio of $10^8$, highlighting its potential for advanced nano-switching applications. These findings underscore the significant impact of metal doping on MoS$_2$'s electronic properties and its promise for future nanoelectronic devices.
\FloatBarrier
	\begin{table}[htbp]
		\centering
		\caption{Average Tunnel resistance (R$_T$) of the considered devices. }
		\label{tab:1}
		\begin{tabular}{cc}
			\hline
			Device        &  R$_T$ ($\Omega$.$\mu$m)  \\ \hline 
			Au-Cd-MoS$_2$ & 2.11E+02	 \\	
                Au-Re-MoS$_2$ & 5.84E+01	 \\
                Au-Ru-MoS$_2$ & 4.82E+00	 \\
			\hline  
		\end{tabular} \\
	\end{table}
\FloatBarrier


\end{document}